\documentclass[aps,pre,twocolumn,showpacs,nofootinbib]{revtex4-1}
\usepackage[dvips]{graphicx}
\usepackage{bm}
\usepackage{mathrsfs}
\usepackage{lipsum}
\usepackage{graphicx,epsfig}
\usepackage{psfrag}
\usepackage{bm}
\usepackage{caption}
\usepackage{subcaption}
\usepackage{epstopdf}
\usepackage{latexsym}
\usepackage{multirow}
\usepackage{amsmath,amssymb,amsfonts}
\usepackage{enumerate}
\usepackage[utf8]{inputenc}
\usepackage{color,soul}
\usepackage{float}
\usepackage[utf8]{inputenc}
\usepackage[utf8]{inputenc}
\usepackage[utf8]{inputenc}
\usepackage{tikz}
\usepackage[utf8]{inputenc}
\allowdisplaybreaks
\begin{document}
\date{\today}

\title{ Effect of geometry on the positioning
	of a single spot in reaction-diffusion systems
}
\author{Sankaran Nampoothiri} \email[email:] {sankaran.n@icts.res.in}%,
%\author{Amal Medhi} \email{amedhi@iisertvm.ac.in}
%\footnote{sankaran.n@icts.res.in}}
\affiliation{ International Centre for Theoretical Sciences, Tata Institute of Fundamental Research, Survey no 151, Shivakote,
	Hesaraghatta Hobli, Bengaluru 560089, India}
\begin{abstract}
We consider the formation of
a single spot (localized solution) in
reaction-diffusion (RD) equation 
on a curved manifold. Specifically,
we study the direction (alignment) of the normal to interface between maxima and minima of concentration in the steady-state on a prolate and on an oblate ellipsoid. We  further analyse the effect of shape asymmetry on $l=1$  eigenmode of the sphere by assuming
a small deformation from the spherical geometry. Our analysis shows that the eigenfunction corresponding to highest eigenvalue  align along the symmetry axis for a prolate ellipsoid, and   perpendicular to the symmetry axis for an oblate ellipsoid. Finally, we compare the direction of variation of the most unstable mode (eigenfunction with highest growth rate) in the system obtained by assuming a small deformation from the sphere and the alignment of interface normal obtain from the numerical simulations.

\end{abstract}

\maketitle
\section{INTRODUCTION}
\paragraph*{}
The emergence of pattern in space
and time is an ubiquitous phenomenon in
nature. Hence its understanding and modeling is of fundamental importance in many fields. After the seminal work of Turing~\cite{turing}, reaction-diffusion systems play a central role in the
mathematical modeling of spatial pattern formation~\cite{murray,biology,vegetation1,vegetation2}.The applicability
of RD equations ranges through many different fields. Recently RD equations have also found its role in understanding the spatial organization of molecules in biological membranes ~\cite{protein}. For example, oscillations of Min protein system in
E.coli cell are modelled using RD equations~\cite{howard,min1,min2}. 

\paragraph*{}
In most of the previous studies RD
equations have been proposed and analyzed
 on flat surfaces to understand the formation of patterns. It is important to note that the features of surface geometry
has been under-appreciated in all these studies. But, the role of surface shape has been
highlighted in many of the recent studies.
%Recently,
%it has become clear that geometry can paly an important role in the cellular
%process. For example, spatial patterning of protein molecules plays an importantpart
For example, the spatial patterning of protein 
molecules are known to be sensitive to cell shape~\cite{thalmeier2016geometry}. Geometry can also play an important role in the formation of complex patterns observed on animal surfaces~\cite{murray}. Thus the shape 
of the surface can be crucial in the
formation of
 wide variety of patterns. 
 %Thus understanding the role of geometry in RD systems can be crucial in many biological process. 
 %These
 %spatial patterns of protein molecules
% plays an importantpart
 %role in cellular process like cell division %and %cellpolarity~\cite{thalmeier2016geometry}. %Hence, the shape of the membrane can play an %important role in orchestrating the cellular %process. Hence the effect 
% of geometry in RD can...
\paragraph*{}
Owing to the importance of understanding RD
on complex geometries,
some of the previous studies have analyzed
the effect of geometry in RD systems~\cite{frank,venkataraman,nucleation,varea,zykov,meandering,ladybeetle,plaza,krause2019influence,nampoothiri2017role}. For example,
the parr-marks formation on fish skins are studied in the work ~\cite{venkataraman} where shape of the skin is 
modeled as growing elliptic cylinder which indicates the importance of surface shape in understanding the patterns observed in nature. %Interestingly,
%the above study also discussed the effect of mean curvature in the formation of
%stripe and spot patterns. 
  Some of the rececnt works about the study on nucleation of RD waves 
on curved surfaces~\cite{nucleation}, spiral waves 
on curved surfaces~\cite{new1} and the effect of geometry on Min-protein dynamics~\cite{walsh2016patterning, wettmann2018effects} again suggest the 
importance of surface shape. These studies suggest that 
the shape of the surface strongly influence the formation of spatial patterns. It is also interesting to note that the importance of 
the spectrum of the Laplace operator on a curved 
surface in RD equation is highlighted in the work~\cite{frank}. 

The localized state (single spot) holds a significant position
 in RD like systems~\cite{localized}. The localized solutions can have important applications in morphogenesis and technologies. One interesting appearance of localized structure is in the RD models of blood coating~\cite{blood}. Another important application of localized solution can come in variuos cellular processes. For example, single spot in RD like equation can play a significant role in cell polarization~\cite{diegmiller2018spherical}. 
 
 In the light of above studies 
 it would be imperative to analyze the interplay between geometry
and the positioning of a single spot in RD. 
In the current work, we have analyzed the role of shape asymmetry of the surface on the positioning of a single spot in RD systems.
To analyze the role of shape asymmetry in RD, we have numerically evolved RD equations
 of Schnakenberg  model on both prolate and oblate ellipsoid. Specifically, we have studied the positions of single spot on both cases by varying one of the parameters in the system. Our analysis suggests that the geometry can act as a cue for the positioning of a single spot in RD systems.
\paragraph*{}
The paper is organized as follows. In Sec. II, we outline the general model of 
RD equation. Then we introduce the Schnakenberg  model of RD and its linear stability analysis on a sphere. We  then consider the formation of a single spot on a sphere using
Schnakenberg model. In Sec. III,
we  analyse the formation of spot on a prolate and on an oblate ellipsoid by solving the RD equations numerically. In Sec. IV,  we carry out a perturbative analysis to understand the effect of deformation on $l=1$ mode by assuming
a small deformation from the spherical geometry. In Sec V, we compare the conclusions from  perturbative analysis and the numerical observations. We summarize our results in Sec. VI.
\section{Model}
In general, the dynamics of RD system on a given curved surface can be modelled 
by  the following set of equations
\begin{subequations}
	\begin{eqnarray}
	\frac{\partial A}{\partial t}&=F_{1}(A,B)+D_{A}\bigtriangleup_{LB} A ,\\                                                     
	\frac{\partial B}{\partial t}&=F_{2}(A,B)+D_{B}\bigtriangleup_{LB} B,
	\label{eq:rd_equation}
	\end{eqnarray}
\end{subequations}
where $A$, $B$ are the concentrations of chemicals,
$F_{1}(A,B),F_{2}(A,B)$
represent the reaction kinetics, $D_{A},D_{B}$ are diffusion coefficients
of the chemicals $A$ and $B$ respectively, and
$\bigtriangleup_{LB}$ is the Laplace-Beltrami operator on curved surface.
The reaction terms in the equation control the degredation and production of chemicals on the surface, and in general, independent of the surface shape. Hence the RD system senses the presence of geometry
through the Laplace-Beltrami operator
of that surface.

\subsection{Schnakenberg model on a sphere}
We restrict to the well-studied Schnakenberg model~\cite{schnakenberg} as 
an example for our studies. The model has the following advantages a) The model has simplest 
kinetics b) The space of parameters 
where the model can exhibit Turing instability is
large and robust. The reaction kinetics of the model
is written as~\cite{schnakenberg}
\begin{eqnarray}
F_{1}(A,B) =k_{1}-k_{2}A+k_{3}A^{2}B,\\
F_{2}(A,B)=k_{4}-k_{3}A^{2}B,
\end{eqnarray}
 where the reaction kinetics $F_{1}$ and $F_{2}$ controls the production and depletion of chemicals $A$ and $B$. To proceed further, we now write the non-dimensional version of the equation as
\begin{subequations}
	\begin{eqnarray}
	\frac{\partial U}{\partial \tau}&=\tilde{\bigtriangleup}_{LB} U+\gamma f(U,V) ,\\                                                      
	\frac{\partial V}{\partial \tau}&=d\tilde{\bigtriangleup}_{LB} V+\gamma g(U,V) ,
	\label{eq:rd_equation}
	\end{eqnarray}
\end{subequations}
where $\tau=\frac{D_{A}t}{a^{2}}$, $\gamma=\frac{a^{2}k_{2}}{D_{A}}$ , $d=\frac{D_{B}}{D_{A}}$, $U=A(k_{3}/k_{2})^{1/2}$ and $V=B(k_{3}/k_{2})^{1/2}$ and the $\tilde{\bigtriangleup}_{LB}$ Laplace-Beltrami operator on a sphere in scaled variable. The reaction kinectics is given by $f(U,V)= (a_{0}-U+U^{2}V)$,   $g(U,V)=
(b_{0}-U^{2}V)$ where $a_{0}=\frac{k_{1}}{k_{2}}(\frac{k_{3}}{k_{2}})^{1/2}$ and
$b_{0}=\frac{k_{4}}{k_{2}}(\frac{k_{3}}{k_{2}})^{1/2}$.
\paragraph*{}
The linear stability analysis about the homogeneous steady state $(U_{0},V_{0})=(a_{0}+b_{0},\frac{b_{0}}{(a_{0}+b_{0})^{2}})$ follows.
A small variation in the homogeneous steady state can be denoted as
%\begin{equation}
%\gamma AW+D\nabla^{2} W
%\end{equation}
\[\delta W=\left (\begin{array}{c}
\delta U-U_{0}\\
\delta V-V_{0}\\
\end{array}\right),\]

which satisfies the linearized equation
\begin{equation}
\frac{\partial~ (\delta W)}{\partial t}=\hat{L}\delta W,
\end{equation}
where
\begin{equation}
\hat{L}=\gamma C+D\nabla^{2},
\end{equation}
\[D=\left (\begin{array}{cc}                      
1&0\\
0&d\\
\end{array}\right),
C=\left (\begin{array}{cc}
\frac{\partial f}{\partial U}&\frac{\partial f}{\partial V}\\
\\
\frac{\partial g}{\partial U}&\frac{\partial g}{\partial V}\\
\end{array}\right)_{U_{0},V_{0}},\]
%\begin{equation}
%\left (\begin{array}{ccc}                      
% \lambda_{0}+\lambda_{2}&0&5\\
% 0&d&7\\
%\end{array}\right)
%\end{equation}

%\begin{equation}
%\begin{bmatrix}
%A&b&g\\c&d&y
%\end{bmatrix} w=0
%\end{equation}

and
\begin{eqnarray*}
	f(U,V)=\gamma(a_{0}-U+
	U^{2}~V),\\
	g(U,V)=\gamma(b_{0}-U^{2}~V)~.
\end{eqnarray*}
\paragraph*{}
The solution to the Eq. (5) can be written as\\
\begin{equation}
\delta W(\theta,\phi,t)=\sum_{l=0}^{\infty}\sum_{m=-l}^{l}C_{l}^{m}e^{\lambda(l) t}
P_{l}^{m}(\cos\theta)e^{im\phi},
\end{equation}
where the constants $C_{l}^{m}$ can be determined from initial conditions. The
eigenvalues $\lambda(l)$ satisfy
\begin{equation}
\lambda^{2}+\lambda[(\frac{l(l+1)}{a^{2}})(1+d)-\gamma(f_{u}+g_{v})]+h(l(l+1))=0.
\end{equation}

Hence the growth rate $\lambda(l)$ corresponding to a particular mode $l$ can be written as
\begin{eqnarray}
\lambda_{\pm}=\frac{-(\frac{l(l+1)}{a^{2}}(1+d)-\gamma(f_{u}+g_{v}))}{2} \nonumber\\\pm  \frac{\sqrt{(\frac{l(l+1)}{a^{2}}(1+d)-\gamma(f_{u}+g_{v}))^{2}-4h(l(l+1))}}{2},
\end{eqnarray}
where $f_{u}=\frac{\partial f}{\partial U}\mid_{U_{0},V_{0}}$, $g_{v}=\frac{\partial g}{\partial V}\mid_{U_{0},V_{0}}$. The $h(l(l+1))$ can be given as
\begin{multline*}
h(l(l+1))=d(l(l+1)/a^{2})^{2}-\gamma(d~f_{u}+g_{v}) ~l(l+1)/a^{2}+\\ \gamma^{2}(f_{u}g_{v}-f_{v}g_{u}).
\end{multline*}
Note that the set of modes having the postive growth rate can lead to spatial inhomogeneity
in  concentration in the steady-state. It is also important to note that the eigenvalues of Laplce-Beltrami
operator is crucial in determining the growth rate for RD systems. 
\paragraph*{}
First we illustrate the formation of single spot on a sphere.
Note that the parameters $(a_{0},b_{0},d,\gamma)$ can control the growth rate of each modes. Hence, these parameters can play a significant role in determining the 
number of spots. In this case, we chose the values of parameters in such a way that $l=1$ mode is unstable and all other modes are stable. We have then numerically solved Eq.(4a,4b) on the surface of a sphere using ~~~FEniCS~\cite{langtangen2016solving}. The single spot obtained on a sphere is shown in the ~~Fig. 1.
\begin{figure}[h!]
	\centering
	\includegraphics[width=0.2\textwidth]{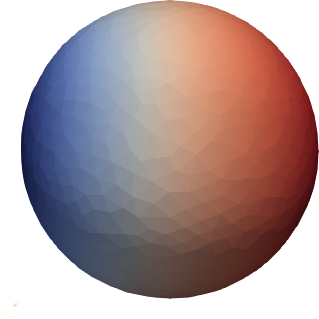}
	\caption{Spot on a sphere. The parameters are $\gamma=8$, $d=10$, $a_{0}=0.1$, $b_{0}=0.9$. The red represents the maxima of the concentration of chemical $U$ and blue represents the minimum. }
\end{figure}
%\subsection{Thomas model on a sphere}

\paragraph*{}

 In the following section we analyse the positioning of a single spot on a prolate and on an oblate ellipsoid by keeping $(a_{0},b_{0},d)$ same as in the case of a sphere and vary the value of parameter $\gamma$. The specific importance of  parameter $\gamma$ and its role in pattern selection is thoroughly discussed~\cite{murray}. Note 
 from Eq. (9) that the changes in the value of $\gamma$ can affect the growth rate of modes. The growth rate of higher modes (modes with lower eigenvalues) can increase as a result of increasing the value of $\gamma$. In other words, the maximum of growth rate can shift towards higher modes  as we increase the value of $\gamma$ as shown in the Fig. 2. 
\begin{figure}[H]
	\centering
	\includegraphics[width=0.5\textwidth]{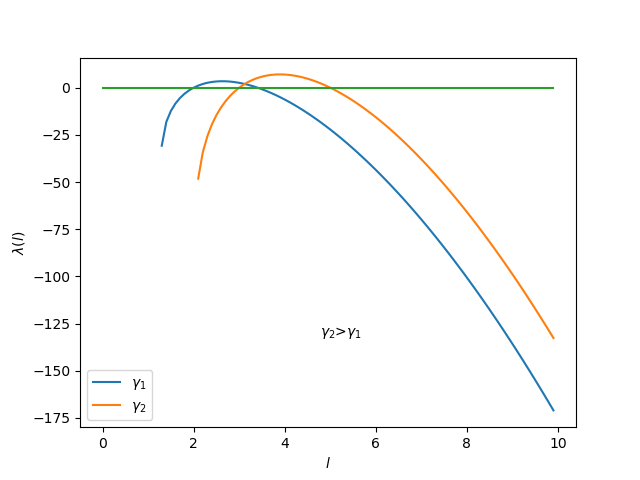}
	\caption{The figure illustrates the effect of $\gamma$ on growth rate $\lambda(l)$. Note that the curve shift towards the left (right) as we decrease (increase) the $\gamma$.  }
\end{figure}
\section{Role of geometry on the positioning 
	of a single spot}
In the previous section we have seen that the role of eigenvalues of Laplace-Beltrami
operator in determining the growth rate  $(\lambda_{\pm})$ of different modes.
It is obvious that  
spectrum (eigenvalues and eigenfunctions) of 
Laplace-Beltrami operator is connected 
to the geometry of the surface. Thus, by controlling the spectrum of Lapalce-Beltrami
operator, the geometry can influence the nature of
the steady-state solutions of RD. Hence, in this
section, we have explored the effect of deformation from the spherical geometry on the localized solution (single spot) in RD systems.
\paragraph*{}
In order to understand the role of geometry, we have numerically solved the Schnakenberg model on both prolate and oblate ellipsoid. Specifically, we have 
studied the positioning of a single spot on both ellipsoids as we vary the parameter $\gamma$. Our numerical simulation shows that the parameter $\gamma$ can play a significant role
in determining the positioning of a single spot .

%In this section we consider Schnakenberg model on prolate and oblate ellipsoids.
%Since the symmetry of the shape has broken in this case, we study about the specific positioning of the single spot on both ellipsoids. We have
%numerically solved the RD equation on both prolate and oblate ellipsoid in order to
%understand the positioning of spot.
%\paragraph*{}
%We then proceed to carry out linear stability
%analysis on both prolate and oblate ellipsoid,
%where we consider a small deviation from sphere. Since we are 
%interested in understanding the positioning 
%of a single spot, we have analyzed 
%the role of deformation 
%on $l=1$ mode. Since the degeneracy of $l=1$ mode is broken by the shape asymmetry the eigenvalues of  different modes $l=1;~~~m=1,-1,0$ can be different.
 
% We have then calculated the correction to eigenvalues 
%corresponding to $l=1;m=1,-1,0$ modes peturbatively. We have computed the growth rate $(\lambda_{\pm})$ for different modes $(l=1;m=1,-1,0)$ and then the correspoding eigen functions at the zeroth 
%order for both prolate and oblate ellipsoids.
%We have then analyzed how the shape deformation affect the growth rate corresponding to different modes.

\subsection{Schnakenberg model on ellipsoid}
In this section we consider the formation of single spot on prolate and oblate ellipsoid
using Schnakenberg model. We now briefly mention the geometrical characterestics of ellipsoids.
The equation of an ellipsoid is
\begin{equation}
\frac{x^{2}+y^{2}}{a^{2}}+\frac{z^{2}}{b^{2}}=1,
%\label{eq:9}
\end{equation}
where the case with $a>b$ is 
called oblate ellipsoid, while the case with $a<b$ is the prolate ellipsoid.
The ellipsoid can be parametrized as
\begin{equation}
X(\theta,\phi) = 
\begin{pmatrix} 
a\sin\theta\cos\phi \\ 
a\sin\theta \sin\phi \\ 
b \cos\theta
\end{pmatrix},
\label{eq:10}
\end{equation}
where $\theta$ and $\phi$ are the coordinates on the surface. 
Note that on an ellipsoid both curvatures are $\theta$ dependent. The Gaussian curvature
of an ellipsoid is positive (see appendix) where the 
curvature varies from $b^{2}/a^{4}$ (at $\theta=0$) to 
$1/b^{2}$ (at $\theta=\pi/2$). Note that the Gauss curvature is maximum at $\theta=0$ and 
minimum at $\theta=\pi/2$ for a prolate  ellipsoid. In the case of an oblate ellipsoid,
the maximum of Gauss curvature occurs at 
$\theta=\pi/2$ and minimum occurs at $\theta=0$.
%\begin{figure}[h!]
	%\centering
	%\includegraphics[width=0.4\textwidth]{gausscurvature.png}
%	\caption{Variation of gauss curvature 
		%on prolate and oblate ellipsoid.
	%}
%\end{figure}
\paragraph*{}
 We have numerically solved 
 the RD equation for Schnakenberg model on both ellipsoids using FEniCS. Initially
 we have considered a homogeneous distribution
 of chemicals on both surfaces. The initial
 condition is then provided by adding random perturbation 
 to the homogeneous steady-state. 
%\begin{figure}[h!]
%	\centering
%	\includegraphics[width=0.4\textwidth]{hhhh.png}
%	\caption{Single spot on a prolate ellipsoid. Top left to right: changes in the position of 
		%maxima of spot as we vary the $\gamma$ and the correspoinding values are $\gamma=8,5.3,4$.We have considered semimajor axis $a=1.1$ and semiminor axis $b=1$. The white line represents the interface. }
%\end{figure}

\paragraph*{}
First, we have solved RD equations on prolate ellipsoid by considereing different values of $\gamma$. The spot obtained in each cases are presnted in the Fig. 3 and ~~~~~~~Fig. 4. To begin with, we have chosen $\gamma=8$ and obtained a single spot in the steady-state where the concentration contains one maxima and one minima as shown in the Fig. 3. In this case, the normal to interface is perpendicular to the axis of symmetry. In other words, concentration is varying perpendicular to the axis of symmetry for
these values of $\gamma$. Here
 the maxima of  concentration is peaked near to the points of minimum Gauss curvature.  We have then considered $\gamma=5.6$ 
and obtained the same positioning of spot as in the previous case as shown in the Fig. 3. 

 We have then carried out the simulation using $\gamma=5.5$ and $\gamma=4$ as shown in the Fig. 4. In both cases, the normal to interface is aligning along the symmetry axis. In other words, the variation of concentration occurs along the symmetry axis in this case.
 Here the maxima of concentration is peaked near to the regions of maximum Gauss curvature . Our simulation shows that concentration can vary along and perpendicular to symmetry axis depending on the parameter $\gamma$.
 \begin{figure}[H]
	\centering
	\includegraphics[width=0.5\textwidth]{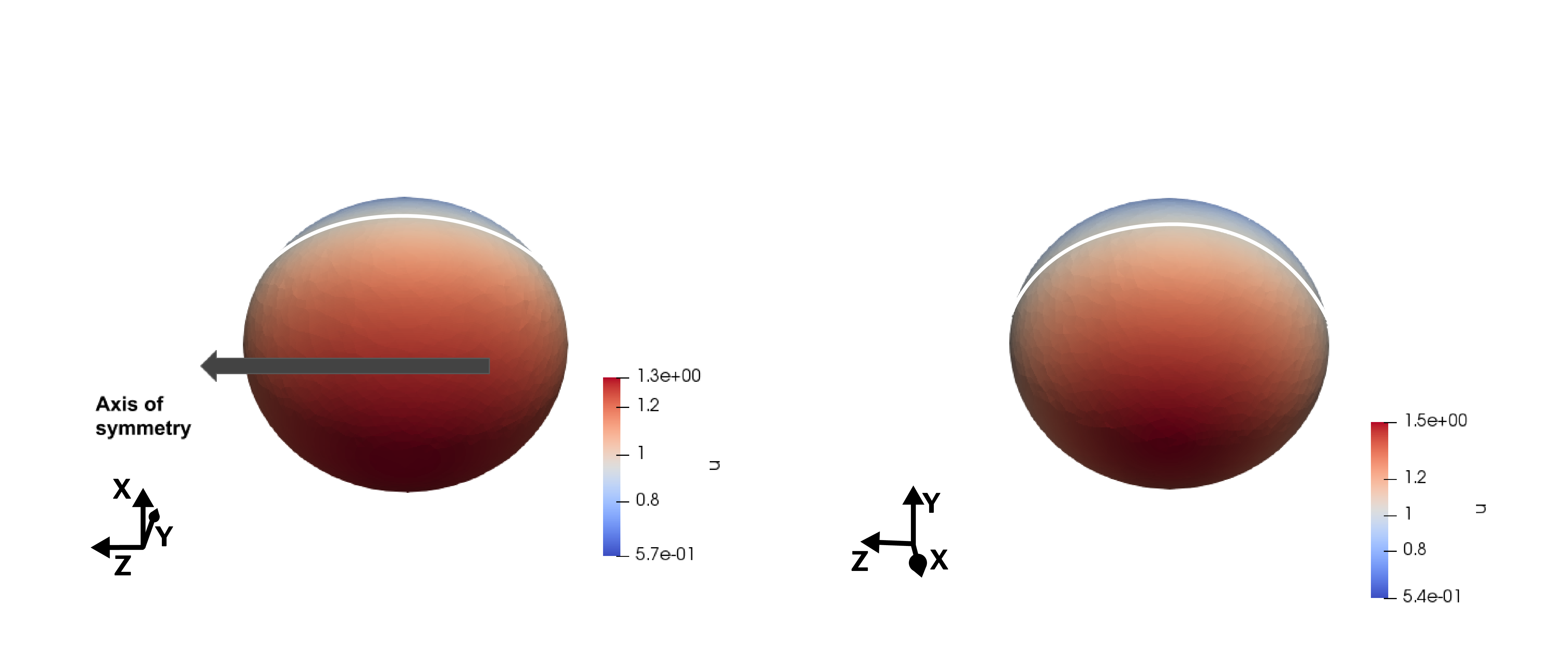}
	\caption{ Left is the single spot on a prolate ellipsoid with $\gamma=8$ and right is the spot obtained for $\gamma=5.6$. The white line represents the interface. The other parameter values are $a_{0}=0.1$, $b_{0}=0.9$, $d=10$. We have chosen semi-major axis $b=1.1$ and semi-minor axis $a=1$. Note that concentration is varying perpendicular to symmetry axis.}
	\end{figure}
\begin{figure}[H]
\centering
	\includegraphics[width=0.5\textwidth]{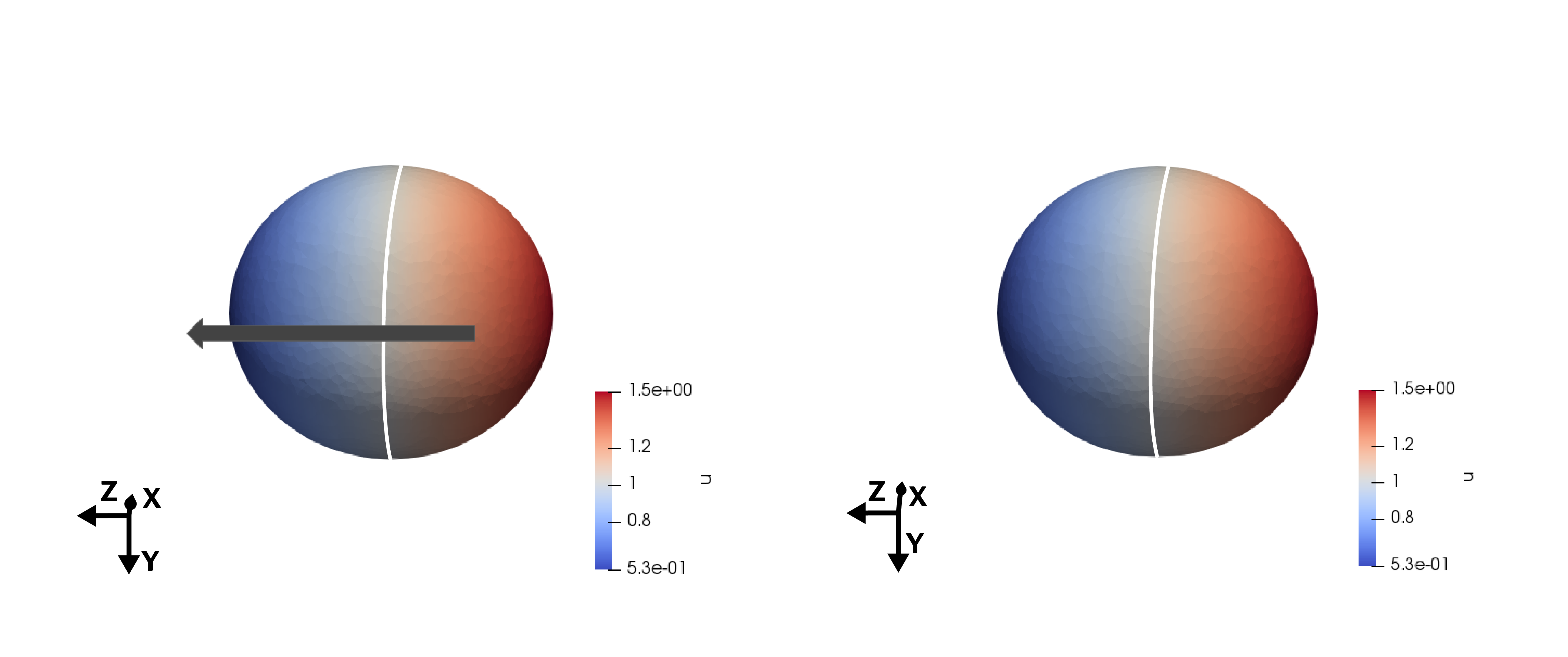}
	\caption{Left is the single spot on a prolate ellipsoid with $\gamma=4$ and right is the spot obtained for $\gamma=5.5$. Note that the interface normal is aligning along symmetry axis.}
\end{figure}

We have then solved the system of RD equations on an oblate ellipsoid and the spot observed for different values of $\gamma$ is shown in the Fig. 5 and Fig. 6. In this case, the paramater $\gamma=8$ and $\gamma=5.5$ result into a patterned state where the normal to interface is aligning along symmetry axis as shown in the Fig. 5. Note that the peak of the maxima is formed around the points of minimum Gauss curvature. 

 We have then considered parameter values $\gamma=5.3$ and $\gamma=4$ in our simulation and obtained a spot as shown
 in the Fig 6.  The interface normal is aligning perpendicular to symmetry axis for both values of $\gamma$. Here the concentration is high around the positions of maximum Gauss curvature. In the case of an oblate ellipsoid also, as similar to prolate ellipsoid, the  concentration can vary along and perpendicular to symmetry axis.
 %\begin{figure}[H]
 	%\centering
 	%\includegraphics[width=0.4\textwidth]{e.png}
 	%\caption{Single spot on an oblate ellipsoid. Top left to right: changes in the position of 
 		%maxima of spot as we vary the $\gamma$ where $\gamma=8, 5.3, 4$. Here $a=1$, $b=1.1$. }
 %\end{figure}
%\subsection{Thomas model on ellipsoid}

	\begin{figure}[H]
		\centering
		\includegraphics[width=0.5\textwidth]{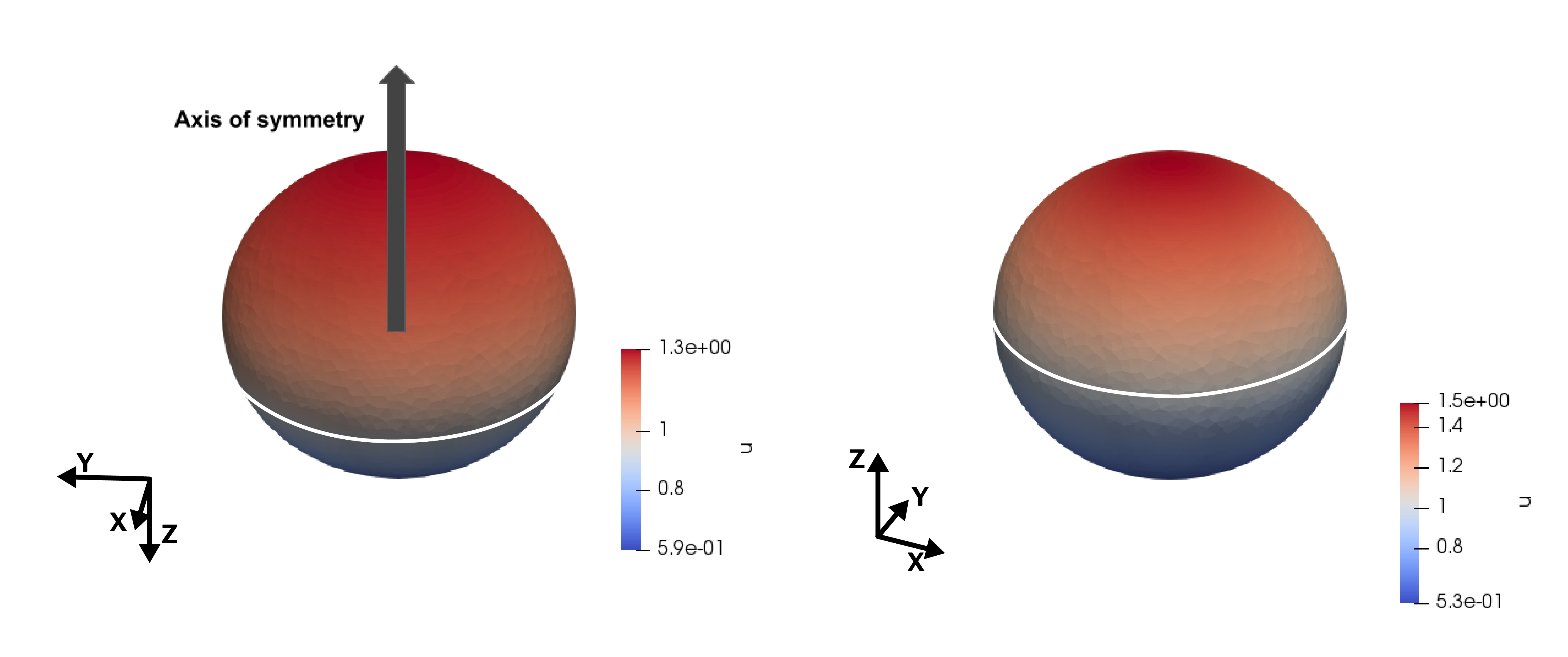}
	\caption{Left is the single spot on an oblate ellipsoid with $\gamma=8$ and right is the spot obtained for $\gamma=5.5$. Note that
	concentration is varying along the symmetry axis. Here $b=1$, $a=1.1$.}
\end{figure}
 \begin{figure}[H]
 	\centering
 \includegraphics[width=0.5\textwidth]{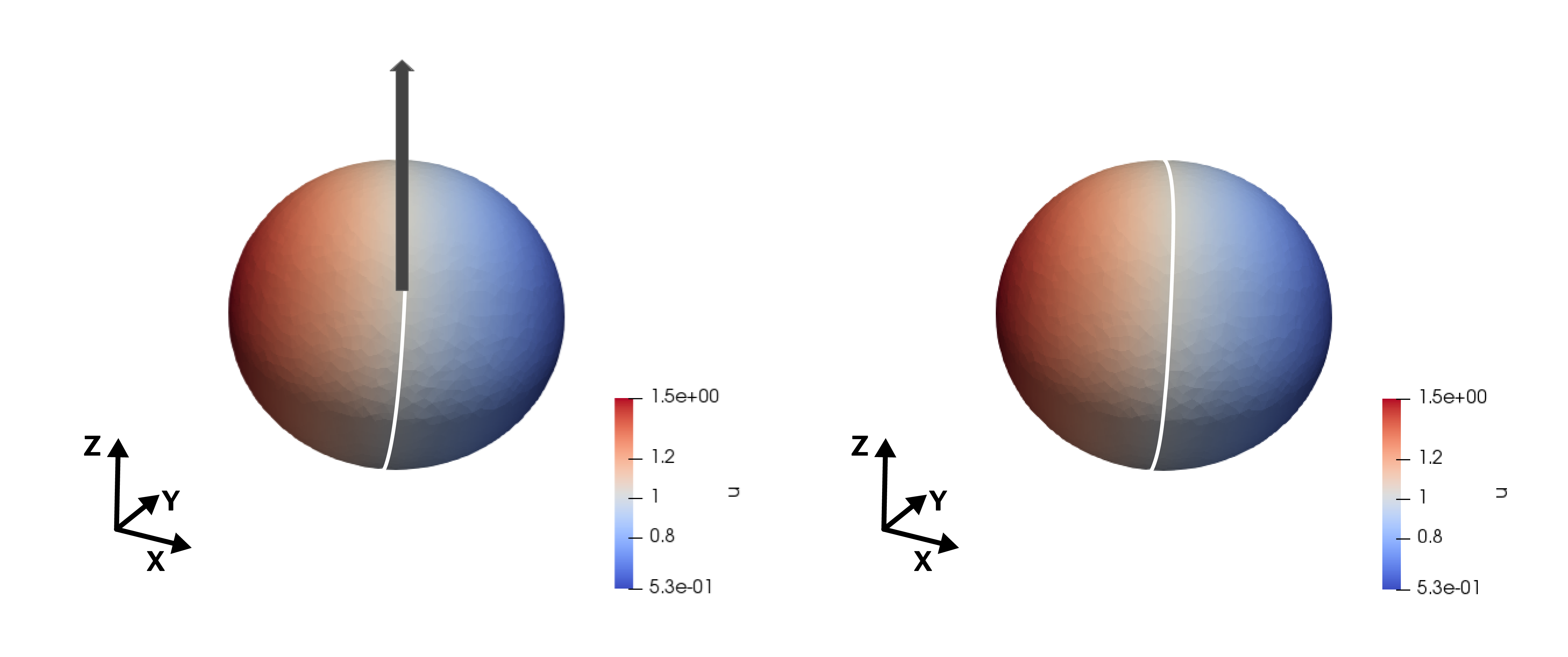}
 	\caption{ Left is the single spot on an oblate ellipsoid with $\gamma=4$ and right is the spot obtained for $\gamma=5.3$. Note 
 	that concentration is varying perpendicular to symmetry axis.}
 \end{figure}
\section{Effect of deformation: prolate and oblate ellipsoid}
%Since the symmetry of the shape has broken in this case, we study about the specific positioning of the single spot on both ellipsoids. We have
%numerically solved the RD equation on both prolate and oblate ellipsoid in order to
%understand the positioning of spot.
%\paragraph*{}
 In this section, we have analyzed 
the role of shape asymmetry
on $l=1$ mode by assuming a small deformation from the spherical geometry. The deformation can remove the degeneracy of $l=1$ mode. In other words, the eigenvalues of  different modes $l=1;~~~m=1,-1,0$ can be different due to shape asymmetry. This result into different growth rates for these modes. Note from Fig. 2 that the modes with lower (higher) eigenvalues can have larger growth rate for high (low) values of $\gamma$. 

In the case of both prolate and oblate ellipsoid,
we have computed perturbatively the correction to eigenvalue of modes $l=1;~~~m=1,-1,0$ to understand which mode has got highest/lowest eigenvalue and also the corresponding eigenfunction to zeroth order.
To summarize, the most unstable mode (eigenfunction with highest growth rate) and its direction of variation (along or perpendicular to symmetry axis) on the surface can be obtained from the perturbative analysis.
%\paragraph*{}
% We summarize the general procedure briefly. First we can write, in general, the growth rate  as
%\begin{eqnarray}
%	\lambda_{\pm}=\frac{-(-\alpha_{l}^{m}(1+d)-\gamma(f_{u}+g_{v}))}{2}\nonumber \\\pm  \frac{\sqrt{(-\alpha_{l}^{m}(1+d)-\gamma(f_{u}+g_{v}))^{2}-4h(\alpha_{l}^{m})}}{2}
%\end{eqnarray}
%where $\alpha_{l}^{m}$ are the eigenvalues of Lapalce-Beltrami operator on the surface of a prolate and oblate ellipsoid.  Since we asumme only a small deformation from the spherical geometry, the eigenvalues $\alpha_{1}^{m=1,-1,0}$ which
%are the corrections to eigenvalue of $l=1$
%mode, can be computed perturbatively. We have then
 %obtained the corresponding eigenfunctions
 %to zeroth order by finding the eigenvectors of the degenaracy matrix in the $l=1$ subspace. To conclude, using the perturbation theory, one can  find out the mode which has got the highest growth rate
%and its direction of variation (alignment) on the surface.

\subsection{ prolate ellipsoid}

First we consider the case of a prolate ellipsoid where we calculate the correction to eigenvalue and 
eigenfunction corresponding to
$l=1;m=1,-1,0$ modes. In oder to obtain the eigenvalues
and eigenfunctions we need to compute 
the form of Laplace-Beltrami operator.
The form of Lapalce-Beltrami operator on any curved surface is given by
$\frac{1}{\sqrt{g}}\partial_{i}\sqrt{g}g^{ij}\partial_{j}$ where $g^{ij}$ is the inverse 
of the metric $g_{ij}$ and $g$ is the determinant of the metric. This can be given as
\begin{multline}
\triangledown^{2}=\frac{1}{a^{2}(\cos^{2}\theta+\frac{b^{2}}{a^{2}}\sin^{2}\theta)}\frac{\partial ^{2}}{\partial \theta^{2}}+\{\frac{\cot\theta}{a^{2}(\cos^{2}\theta+\frac{b^{2}}{a^{2}}\sin^{2}\theta}+\\\frac{(1-\frac{b^{2}}{a^{2}})\sin2\theta}{2a^{2}(\cos^{2}\theta+\frac{b^{2}}{a^{2}}\sin^{2}\theta)^{2}}\}\frac{\partial}{\partial \theta}+\frac{1}{a^{2}\sin^{2}\theta}\frac{\partial ^{2}}{\partial \phi^{2}}.
\end{multline}
 The above form of $\bigtriangledown^{2}$ 
for a small deformation from the spherical geometry can be written as 
\begin{equation}
\bigtriangledown^{2}=\bigtriangledown^{2}_{\text{sphere}}+\hat{A},
\end{equation}
where $\hat{A}$ is given as 
\begin{equation}
\hat{A}=-\frac{2\epsilon}{a^{3}}
\sin^{2}\theta \frac{\partial^{2}}{\partial \theta^{2}}-\frac{2\epsilon}{a^{3}}\sin2\theta \frac{\partial }{\partial\theta}.
\end{equation}
\paragraph*{}
Next, for $l=1$ case where we have a 3-fold degeracy, we compute the $3\times3$ matrix of the perturbation $\hat{A}$ which is given by
\begin{equation}
\begin{bmatrix}
\langle Y_{1}^{1}\mid\hat{A}\mid Y^{1}_{1}\rangle & 	\langle Y_{1}^{1}\mid\hat{A}\mid Y^{0}_{1}\rangle & 	\langle Y_{1}^{1}\mid\hat{A}\mid Y^{-1}_{1}\rangle  \\
\langle Y_{1}^{0}\mid\hat{A}\mid Y^{1}_{1}\rangle & 	\langle Y_{1}^{0}\mid\hat{A}\mid Y^{0}_{1}\rangle & 	\langle Y_{1}^{0}\mid\hat{A}\mid Y^{-1}_{1}\rangle  \\
\langle Y_{1}^{-1}\mid\hat{A}\mid Y^{1}_{1}\rangle & 	\langle Y_{1}^{-1}\mid\hat{A}\mid Y^{0}_{1}\rangle & 	\langle Y_{1}^{-1}\mid\hat{A}\mid Y^{-1}_{1}\rangle
\end{bmatrix}.
\end{equation}
We consider the following form of spherical harmonics for calculating the above matrix elements.
\begin{eqnarray*}
	Y_{1}^{1}(\theta,\phi)&=&\frac{-1}{2}\sqrt{3/2\pi}\sin\theta e^{i\phi},\\
	Y_{1}^{0}&=&\frac{1}{2}\sqrt{3/\pi}\cos\theta,\\
	Y_{1}^{-1}&=&\frac{1}{2}\sqrt{3/2\pi}\sin\theta e^{-i\phi}.
\end{eqnarray*}
Because of the orthogonality relation we need to calculate only the integrals in the diagonal terms
of the perturbation matrix. These are evaluated to be
\begin{eqnarray*}
	\langle Y_{1}^{1}\mid \hat{A}\mid Y_{1}^{1}\rangle&=&\frac{4\epsilon}{5a^{3}},\\
	\langle Y_{1}^{-1}\mid \hat{A}\mid Y_{1}^{-1}\rangle&=&\frac{4\epsilon}{5a^{3}},\\
	\langle Y_{1}^{0}\mid \hat{A} \mid Y_{1}^{0}\rangle
	&=&\frac{12\epsilon}{5a^{3}}.\\
\end{eqnarray*}
The perturbation matrix can now be explicitly written as
\begin{equation} 
\begin{bmatrix}
4\epsilon/5a^{3} & 0 & 0 \\
0 & 12\epsilon/5a^{3} &0 \\
0 & 0 & 4\epsilon/5a^{3}
\end{bmatrix}.
\end{equation}
We can now write the correction to eigenvalue for $l=1$ mode due to a small deformation from spherical geometry as 
\begin{eqnarray}
\alpha_{1}^{1}&=&-2/a^{2}+4\epsilon/5a^{3},\\
\alpha_{1}^{0}&=&-2/a^{2}+12\epsilon/5a^{3},\\
\alpha_{1}^{-1} &=&-2/a^{2}+4\epsilon/5a^{3},
\end{eqnarray}
where $\alpha_{1}^{1,0,-1}$ are the new
eigenvalues calculated upto $\mathcal O(\epsilon)$. Note from above expression that the eigenvalue $\alpha_{1}^{0}$ is higher compared to other eigenvalues.
The three-fold degenaracy of $l=1$ mode is lifted to two-fold degenaracy due to deformation. Now we need to calculate the eigen functions corresponding to these eigenvalues. 
\paragraph*{}
The eigenvectors of the perturbation matrix  are given by 
\begin{equation} 
\mid I \rangle=\begin{bmatrix}
1 \\
0 \\
0
\end{bmatrix},
\mid II\rangle=\begin{bmatrix}
0 \\
1 \\
0
\end{bmatrix},
\mid III \rangle=\begin{bmatrix}
0 \\
0 \\
1
\end{bmatrix}.
\end{equation}
 Now one can write the  eigenvectors to zeroth order corresponding to $\alpha_{1}^{1},\alpha_{1}^{0}$ and $\alpha_{1}^{-1}$ as
\begin{eqnarray}
\mid \psi_{1~1} \rangle &=&Y_{1}^{1},\\
\mid \psi_{1~2} \rangle&=&Y_{1}^{0},\\
\mid \psi_{1~3} \rangle&=&Y_{1}^{-1}.
\end{eqnarray}
Note that the eigenfunction correponding to highest eigen value $\alpha_{1}^{0}$ is given by $Y_{1}^{0}$ which is varying along the axis of symmetry for a prolate ellipsoid. The eigenfunction ($Y_{1}^{-1},Y_{1}^{1}$) corresponding to lowest eigenvalue is varying perpendicular to symmetry axis.

%\begin{eqnarray}
%\alpha_{1}^{-1}&=&-l(l+1)/a^{2}+4\epsilon/5a^{3},\\
%\alpha_{1}^{0}&=&-l(l+1)/a^{2}+12\epsilon/5a^{3},\\
%\alpha_{1}^{-1}&=&-l(l+1)/a^{2}+4\epsilon/5a^{3}.
%\end{eqnarray}

\paragraph*{}
Note that the growth rate of modes are different as a result of removing the degeneracy in the eigenvalues due to deformation. We can write the growth rate $\lambda_{\pm}$ corresponding to the different modes
$Y_{1}^{0}$ and $Y_{1}^{1}$ by following the Eq. (9) as

\begin{eqnarray}
	\lambda^{0}_{1\pm}=\frac{-(-\alpha_{1}^{0}(1+d)-\gamma(f_{u}+g_{v}))}{2} \nonumber\\\pm  \frac{\sqrt{(-\alpha_{1}^{0}(1+d)-\gamma(f_{u}+g_{v}))^{2}-4h(\alpha_{1}^{0})}}{2},\\
\lambda^{1}_{1\pm}=\frac{-(-\alpha_{1}^{1}(1+d)-\gamma(f_{u}+g_{v}))}{2}\nonumber \\\pm  \frac{\sqrt{(-\alpha_{1}^{1}(1+d)-\gamma(f_{u}+g_{v}))^{2}-4h(\alpha_{1}^{1})}}{2},	
\end{eqnarray}
where $	\lambda^{0}_{1\pm}$ and $\lambda^{1}_{1\pm}$ are the growth rates corresponding to modes $Y_{1}^{0}$ and $Y_{1}^{1}$ . Note from Fig. 2  that the higher modes (modes with lower eigenvalues) can have largest growth rate
for high values of $\gamma$. The lower modes can become more unstable for low values of $\gamma$. 
\paragraph*{}
In the case of a prolate ellipsoid, we have seen from the Eq. (17, 18, 19) that the eigenvalue of mode $Y_{1}^{1}$ is lower compared to $Y_{1}^{0}$.  Hence the growth rate $\lambda_{1+}^{1}$ of the mode $Y_{1} ^{1}$ with lower eigenvalue can be larger compared to the growth rate of $Y_{1}^{0}$ for high values of $\gamma$. Note that $Y_{1}^{1}$ is varying perpendicular to the axis of symmetry. The mode $Y_{1}^{0}$ with higher eigenvalue can become more unstable $(\lambda_{1+}^{0}>\lambda_{1+}^{1})$ for lower values of $\gamma$. The mode $Y_{1}^{0}$ is varying along the symmetry axis. The schematic illustration of the effect of $\gamma$ on the growth rate of modes is shown in the Fig. 7.
\begin{figure}[h!]
	\centering
	\includegraphics[width=0.4\textwidth]{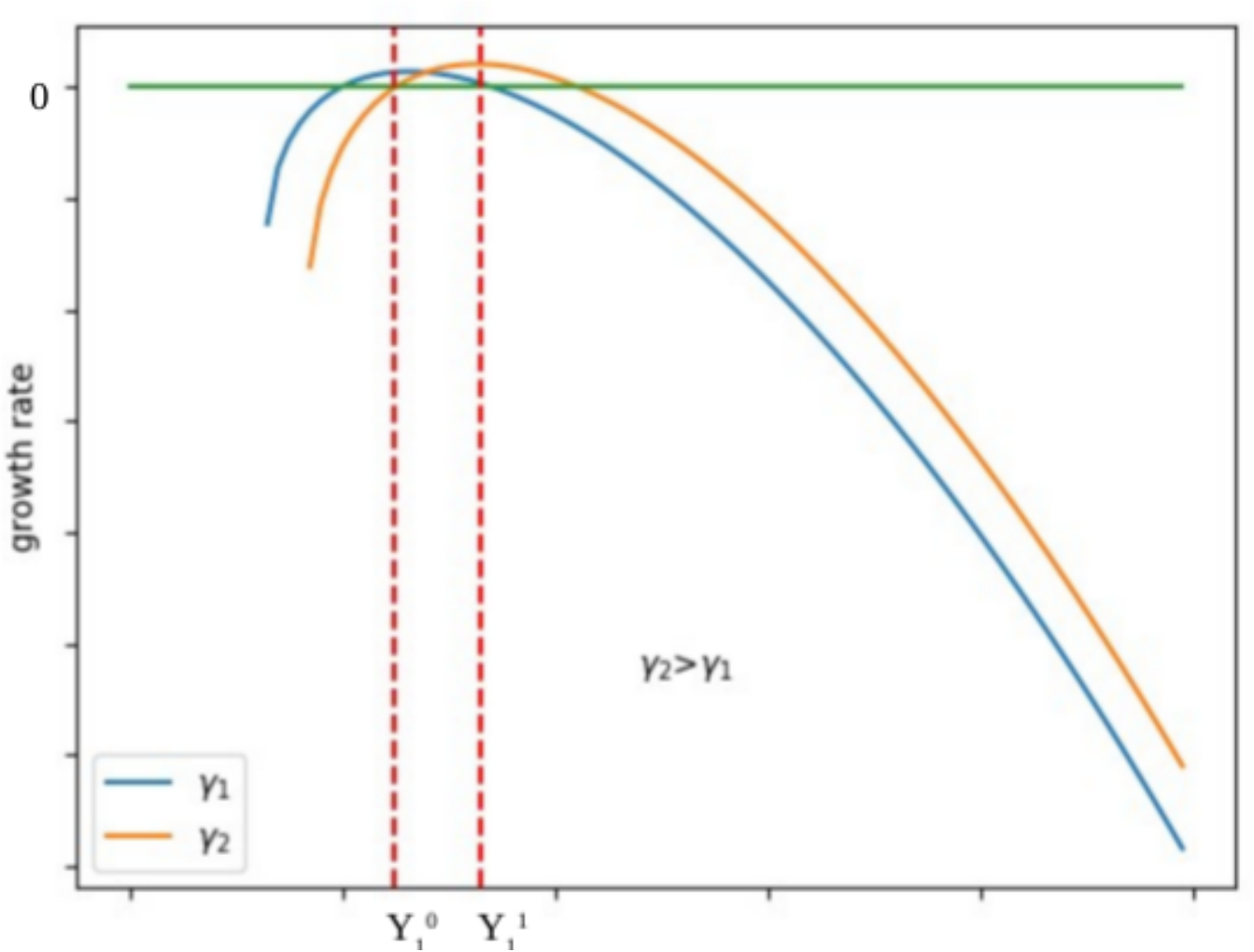}
	\caption{Schematic illustration of the effect of $\gamma$ on growth rate of the modes
$Y_{1}^{0}$ and $Y_{1}^{1}$. Note that
the mode $Y_{1 }^{0}$ can become more unstable as we decrease the $\gamma$ from $\gamma_{2}$ to $\gamma_{1}$. $Y_{1}^{0}$ vary along symmetry axis and $Y_{1}^{1}$ vary perpendicular to symmetry axis.}
\end{figure}
\subsection{ oblate ellipsoid}
Here, similar to the analysis carried out in the case of a prolate ellipsoid, we calculate the growth rate corresponding to different modes. We can write the form of $\bigtriangledown^{2}$ for a small deformation from the spherical geometry as
\begin{equation}
\bigtriangledown^{2}=\bigtriangledown^{2}_{\text{sphere}}+\hat{A},
\end{equation}
where $\hat{A}$ is given by
\begin{equation}
\hat{A}=\frac{2\epsilon}{a^{3}}
\sin^{2}\theta \frac{\partial^{2}}{\partial \theta^{2}}+\frac{2\epsilon}{a^{3}}\sin2\theta \frac{\partial }{\partial\theta}.
\end{equation}
We can now calculate the correction eigenvalues for $l=1$ mode using the perturbation theory and can be given as
\begin{eqnarray}
\alpha_{1}^{-1}&=&-2/a^{2}-4\epsilon/5a^{3},\\
\alpha_{1}^{0}&=&-2/a^{2}-12\epsilon/5a^{3},\\
\alpha_{1}^{-1}&=&-2/a^{2}-4\epsilon/5a^{3}.
\end{eqnarray}
In the case of an oblate ellipsoid, the eigenfunction correponding to higher eigenvalue $\alpha_{1}^{1}$ or $\alpha_{1}^{1}$ is given by $Y_{1}^{1}$ or $Y_{1}^{-1}$. The eigenvalue of the mode $Y_{1}^{0}$ is lower compared to eigenvalue of the mode $Y_{1}^{1}$ in the case of an oblate ellipsoid.

 In the case of an oblate ellipsoid, the growth rate of $Y_{1} ^{0}$ with lower eigenvalue can be higher compared to the growth rate of $Y_{1}^{1}$ for high $\gamma$ values. The mode $Y_{1}^{1}$ with higher eigenvalue can become more unstable for low values of $\gamma$. The schematic illustration of the effect of $\gamma$ on the growth rate of modes is shown in the Fig 8.
 \begin{figure}[h!]
 	\centering
 	\includegraphics[width=0.4\textwidth]{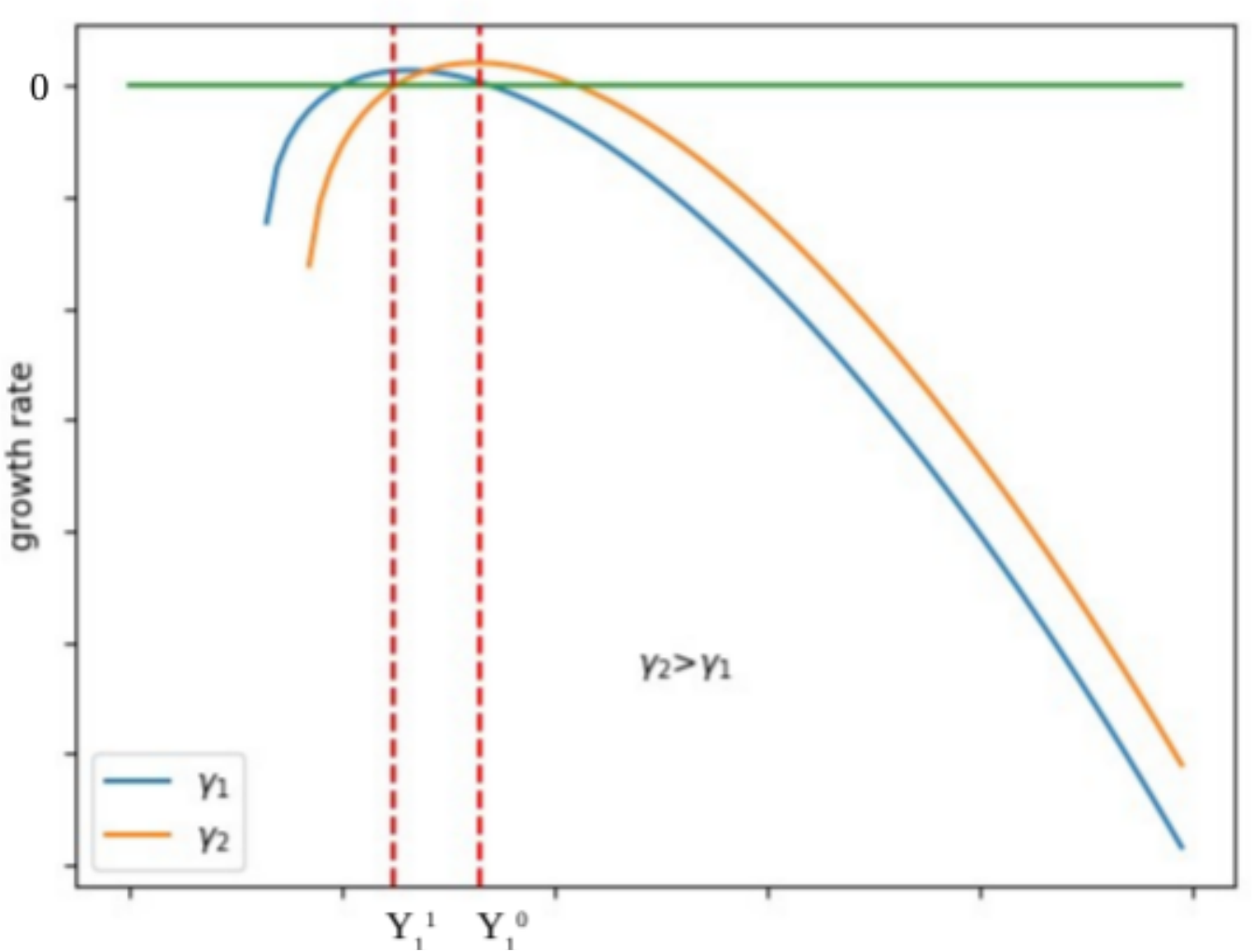}
 	\caption{Schematic illustration of the effect of $\gamma$ on growth rate of the modes
 		$Y_{1}^{0}$ and $Y_{1}^{1}$. Note that
 	the mode $Y_{1 }^{1}$ can become more unstable as we decrease the $\gamma$ from $\gamma_{2}$ to $\gamma_{1}$.}
 \end{figure}
 \section{Direction of variation of most unstable mode and numerical observations: A comparison}
 %\subsection{Prolate ellipsoid}
 \paragraph*{}
 In this section, we compare the direction of variation of the most unstable mode obtained by the perturbative analysis and the direction of interface normal observed in our numerical simulations. The normal to interface is aligning perpendicular to symmetry axis for $\gamma=8$ and $\gamma=5.6$ and parallel to symmetry axis for $\gamma=5.5$ and
 $\gamma=4$ in the case of a prolate ellipsoid.
   Thus our numerical simulations shows that
   concentration can vary perpendicular (parallel) to symmetry axis for 
   high (low) values of $\gamma$ . 
   
    The perturbative analysis shows that the mode  $Y_{1}^{1}$ ($Y_{1}^{0}$) can be more unstable for high (low) values of $\gamma$ 
 as schematically shown in the Fig. 7 in the case of a prolate ellipsoid. Hence, our analysis suggests that the
 mode varying perpendicular (parallel) to symmetry axis can be more unstable for the   
 high (low) values of $\gamma$. Note 
  the similarity between the directions of variation of concentration observed in the simulations and the directions of variation of most unstable mode obtained by the perturbative analysis for high (low)
 values of $\gamma$.

 %\subsection{Oblate ellipsoid}
 \paragraph*{}
 The interface normal is aligning along the symmetry axis for $\gamma=8$ and $\gamma=5.5$
 and perpendicular to symmetry axis for
 $\gamma=5.3$ and $\gamma=4$ in the case of an oblate ellipsoid. The perturbative analysis shows that the mode  varying  parallel (pependicular) can be more unstable for high (low) values of $\gamma$ as 
 schematically illustrated in the Fig. 8. The analysis again indicates the similarity between the directions of interface normal observed 
 in the simulations and the directions of most unstable mode.
\section{Summary}
\paragraph*{}
To sum up, we have studied the role of shape asymmetry on the positioning of a single spot
using Schnakenberg model on both prolate and oblate ellipsoid. In the case of a prolate ellipsoid, the normal to interface is aligning perpendicular to symmetry axis for $\gamma=8$ and $\gamma=5.6$. For values of $\gamma=5.5$ and $\gamma=4$, the interface normal is aligning along the symmetry axis.
In the case of an oblate ellipsoid, for $\gamma=8$ and $\gamma=5.5$, the normal to interface is aligning along the symmetry axis.   The normal to interface is aligning pependicular to symmetry axis for $\gamma=5.3$ and $\gamma=4$. 
In both prolate and oblate ellipsoid, the concentration can vary along and perpendicular
to the symmetry axis depending on the parameter value $\gamma$. 
%\paragraph*{}
%We have further carried out the linear stability analysis for both prolate and oblate ellipsoid, to understand the growth rate of modes. Our analysis shows that the high value of $\gamma$ favours the  mode $Y_{1}^{-1}$  to have the highest growth rate in the case of a prolate ellipsoid and  the mode is varying perpendicular to the axis of symmetry. The lower value of $\gamma$  Specifically, high value of $\gamma$ favour the mode $Y_{1}^{-1}$ to have the highest growth rate initilally on a prolate ellipsoid. This eigenfunction vary perpendicular to symmetry axis. Then the lower mode can also become unstable as a result in the decrese of $\gamma$ as shown in the Fig. Further decrese in the $\gamma$ can favour the lowest mode $Y_{1}^{1}$ to have the highest growth rate in the case of prolate ellipsoid. This eigenfunction align along the symmetry axis for a prolate ellipsoid.
\paragraph*{}
We have analysed the effect of shape asymmetry
on $l=1$ mode by assuming a small deformation from 
the spherical geometry. Our analysis shows that
the  mode
$Y_{1}^{1}$ ($Y_{1}^{0}$) can become more 
unstable for high (low) values of $\gamma$
for a prolate ellipsoid. In the case of an 
oblate ellipsoid, the  mode
$Y_{1}^{0}$ ($Y_{1}^{1}$) can become more 
unstable for high (low) values of $\gamma$.
\paragraph*{} 

    We have then compared the direction of variation of concentration obtained in the numerical simulation with the direction of variation of the most unstable mode obtained by the perturbative analysis. The concentration
    can vary perpendicular (parallel) to symmetry axis as we move along high (low) values of $\gamma$ for a prolate ellipsoid.
    The mode $Y_{1}^{1}$ ($Y_{1}^{0})$ 
   can be more unstable for high (low) values 
   of $\gamma$ as schematically illustrated in Fig. 7 in the case of a prolate ellipsoid. 
   \paragraph*{}
In the case of an oblate ellipsoid, concentration
can vary parallel (perpendicular) to symmetry axis as we move along high (low) values of $\gamma$. In this case, the mode $Y_{1}^{0}$ ($Y_{1}^{1}$)
can become more unstable for high (low)
values of $\gamma$ as schematically shown in the Fig. 8. Thus, in the case of both prolate and oblate
ellipsoid, we have observed a similarity between the directions of interface normal
observed in the simulations and the directions of most unstable mode obtained from the perturbative analysis.

\paragraph*{}
  The analysis presented in the work can be extended to understand the role of geometry in any RD models like BVAM model~\cite{varea} and other models~\cite{murray,biology}. Another important application of this work can come in understanding the role of geometry in various cellular process. Many important processes in cell and developmental biology are controlled by the spatial distribution of proteins~\cite{halatek2018self} where
   the effect of geometry can be significant~\cite{curvature}. Note that RD like equations play a crucial role~\cite{goryachev2008dynamics,otsuji2007mass,howard,min1,min2} in understanding these spatial distribution of proteins. Hence the analysis presented here can be incorporated into 
   above studies which may provide useful insights about the role of cell geometry in various cellular processes.

    The localized solution (single spot) of RD systems can play a significant role in determining the positioning of plane of division in cell division processes~\cite{diegmiller2018spherical}.  The current study hints that identifying
    the possible directions of variation of eigenfunctions can lead to a
    model independent (neglect the details of reaction kinetics) understanding of the positioning 
    of a single spot on arbitarly shaped surfaces. Thus the analysis presented here may be useful to  give insights about the possible planes of division without knowing the details of reaction kinetics.   \section*{Acknowledgements}
   I thank Vijaykumar Krishnamurthy 
   for many useful discussions about the FEniCS. I also
   thank Vinayak Jagadish for careful 
   reading of the manuscript and valuable suggestions.
\appendix
\section{ Laplace operator and perturbative calculations}
Here we calculate the correction to eigenvalues
and eigenfunction of $l=1$ mode due to deformation from spherical geometry. First we consider the case of a prolate ellipsoid.
\paragraph*{}
The ellipsoid can be parametrized as
\begin{eqnarray}
x&=&a \sin\theta \cos\phi,\\
y&=&a \sin\theta \sin\phi,\\
z&=&b\cos\theta,
\end{eqnarray}
where the range of $\phi$ and $\theta$ is given by
$0\leqslant \phi\leqslant 2\pi$ and $0\leqslant \theta\leqslant \pi$. The vector $\vec{X}$ is given by
\begin{equation}
\vec{X}=a \sin\theta \cos\phi \hat{i}+a \sin\theta \sin\phi \hat{j}+b\cos\theta \hat{k}.
\end{equation}
The metric $g_{\theta\theta}$ and 
$g_{\phi\phi}$ is given by
\begin{eqnarray*}
	g_{\theta\theta}&=&a^{2}(\cos^{2}\theta+\frac{b^{2}}{a^{2}}\sin^{2}\theta),\\
	g_{\phi\phi}&=&a^{2}\sin^{2}\theta,\\
	g_{\theta\phi}&=&g_{\phi\theta}=0.
\end{eqnarray*}
The form of Lapalce-Beltrami operator is given by
$\frac{1}{\sqrt{g}}\partial_{i}\sqrt{g}g^{ij}\partial_{j}$. This can be explicitily calculated as
\begin{multline*}
\triangledown^{2}=\frac{1}{a^{2}(\cos^{2}\theta+\frac{b^{2}}{a^{2}}\sin^{2}\theta)}\frac{\partial ^{2}}{\partial \theta^{2}}+\{\frac{\cot\theta}{a^{2}(\cos^{2}\theta+\frac{b^{2}}{a^{2}}\sin^{2}\theta}+\\\frac{(1-\frac{b^{2}}{a^{2}})\sin2\theta}{2a^{2}(\cos^{2}\theta+\frac{b^{2}}{a^{2}}\sin^{2}\theta)^{2}}\}\frac{\partial}{\partial \theta}+\frac{1}{a^{2}\sin^{2}\theta}\frac{\partial ^{2}}{\partial \phi^{2}}.
\end{multline*}
Note that when $b=a$ the form of the Laplace operator reduces to sphere as expected. Now consider a small deformation of the form $b=a+\epsilon$. We can write the term $
(\cos^{2}\theta+\frac{b^{2}}{a^{2}}\sin^{2}\theta)$ as $(1+\frac{2\epsilon}{a}\sin^{2}\theta)$ by neglecting $\mathcal O(\epsilon^{2})$ term.
Now we can write $\bigtriangledown^{2}$ to 
$\mathcal O(\epsilon)$ as 
\begin{eqnarray*}
\bigtriangledown^{2}=\frac{1}{a^{2}}\frac{\partial ^{2}}{\partial \theta^{2}}+\frac{\cot\theta}{a^{2}}
\frac{\partial }{\partial \theta}+\frac{1}{a^{2}\sin^{2}\theta}\frac{\partial ^{2}}{\partial \phi^{2}}-\\\frac{2\epsilon}{a^{3}}
\sin^{2}\theta \frac{\partial^{2}}{\partial \theta^{2}}-\frac{2\epsilon}{a^{3}}\sin2\theta \frac{\partial }{\partial\theta}.
\end{eqnarray*}
The above form of $\bigtriangledown^{2}$ 
can be written as 
\begin{equation}
\bigtriangledown^{2}=\bigtriangledown^{2}_{sphere}+\hat{A},
\end{equation}
where $\hat{A}$ is given as 
\begin{equation}
\hat{A}=-\frac{2\epsilon}{a^{3}}
\sin^{2}\theta \frac{\partial^{2}}{\partial \theta^{2}}-\frac{2\epsilon}{a^{3}}\sin2\theta \frac{\partial }{\partial\theta}.
\end{equation}
The elements in the peturbation matrix $\hat{A}$ is given by
\begin{multline*}
\langle Y_{1}^{1}\mid \hat{A}\mid Y_{1}^{1}\rangle =2\pi \int_{0}^{\pi}\frac{-1}{2}\sqrt{3/2\pi}\sin\theta
\\ \{\frac{2\epsilon}{a^{3}}\sin^{2}\theta\frac{\partial^{2}(\sqrt{3/2 \pi}\sin\theta)}{\partial \theta^{2}}+\frac{2\epsilon}{a^{3}}\sin2\theta\frac{\partial(\sqrt{3/2 \pi}\sin\theta)}{\partial \theta}\}\sin\theta ~d\theta\\=\frac{4\epsilon}{5a^{3}}.\\
\langle Y_{1}^{-1}\mid \hat{A}\mid Y_{1}^{-1}\rangle=\frac{4\epsilon}{5a^{3}}.\\
\langle Y_{1}^{0}\mid \hat{A} \mid Y_{1}^{0}\rangle=
2\pi \int_{0}^{\pi}\frac{-1}{2}\sqrt{3/2\pi}\cos\theta
\\\{\frac{2\epsilon}{a^{3}}\sin^{2}\theta\frac{\partial^{2}(\sqrt{3/ \pi}\cos\theta)}{\partial \theta^{2}}\\+\frac{2\epsilon}{a^{3}}\sin2\theta\frac{\partial(\sqrt{3/ \pi}\cos\theta)}{\partial \theta}\}\sin\theta ~d\theta=\frac{12\epsilon}{5a^{3}}.\\
\end{multline*}
Similar procedure can be done for an oblate ellipsoid also.
\section{Ellipsoid}
The ellipsoid can represented as
\begin{equation}
\frac{x^{2}+y^{2}}{a^{2}}+\frac{z^{2}}{b^{2}}=1,
%\label{eq:9}
\end{equation}
where the case with $a>b$ is 
called oblate spheroid, while the case with $a<b$ is prolate spheroid.
The ellipsoid can be parametrized as
\begin{equation}
X(\theta,\phi) = 
\begin{pmatrix} 
a\sin\theta\cos\phi \\ 
a\sin\theta \sin\phi \\ 
b \cos\theta
\end{pmatrix},
\label{eq:10}
\end{equation}
where $\theta$ and $\phi$ are the coordinates on the surface. 
Using the above parametrization we read intrinsic and extrinsic quantities
related to curvature as
\begin{eqnarray*}
	g_{\theta\theta}&=&a^{2}(\cos^{2}\theta+\frac{b^{2}}{a^{2}}\sin^{2}\theta)
	,\;g_{\phi\phi}=a^{2}\sin^{2}\theta,\;g_{\theta \phi}=g_{\phi \theta}=0,\\
	\kappa_{\theta\theta}&=&\frac{b}{(\cos^{2}\theta+\frac{b^{2}}{a^{2}}\sin^{2}\theta)^{1/2}},\;
	\kappa_{\phi\phi}=\frac{b\sin^{2}\theta}{(\cos^{2}\theta+\frac{b^{2}}{a^{2}}\sin^{2}\theta)^{1/2}},\\
	\kappa_{\theta \phi}&=&\kappa_{\phi \theta}=0,
	\label{eq:14}
\end{eqnarray*}
and then Gauss and mean curvature on an ellipsoid is given by
\begin{eqnarray*}
	K = \frac{b^{2}}{a^{4}(\cos^{2}\theta+\frac{b^{2}}{a^{2}}\sin^{2}\theta)^{2}},\\
	H=b\frac{1+(\cos^{2}\theta+\frac{b^{2}}{a^{2}}\sin^{2}\theta)}{2 a^{2}
		(\cos^{2}\theta+\frac{b^{2}}{a^{2}}\sin^{2}\theta)^{3/2}}.
	\label{eq:15}
\end{eqnarray*}

\bibliography{ref}
\end{document}